\documentclass{article}
\usepackage[margin=0.75in]{geometry}
\usepackage{url}
\usepackage[utf8]{inputenc}
\usepackage[english]{babel}
\usepackage{appendix}
\usepackage{graphicx}

\usepackage{biblatex}
\title{Undecidability Results and Their Relevance in Modern Music Making}
\date{September 2023}
\author{Halley Young, \texttt{halleyy@seas.upenn.edu}}
\begin{document}
\maketitle


\begin{abstract}
This paper delves into the intersection of computational theory and music, examining the concept of undecidability and its significant, yet overlooked, implications within the realm of modern music composition and production. It posits that undecidability, a principle traditionally associated with theoretical computer science, extends its relevance to the music industry. The study adopts a multidimensional approach, focusing on five key areas: (1) the Turing completeness of Ableton, a widely used digital audio workstation, (2) the undecidability of satisfiability in sound creation utilizing an array of effects, (3) the undecidability of constraints on polymeters in musical compositions, (4) the undecidability of satisfiability in just intonation harmony constraints, and (5) the undecidability of "new ordering systems". In addition to providing theoretical proof for these assertions, the paper elucidates the practical relevance of these concepts for practitioners outside the field of theoretical computer science. The ultimate aim is to foster a new understanding of undecidability in music, highlighting its broader applicability and potential to influence contemporary computer-assisted (and traditional) music making.
\end{abstract}

\section{Introduction - Brief overview of the paper's goals}
The primary objective of this paper is to explore the concept of undecidability within the context of modern music making. By examining various aspects of music production and theory, we aim to demonstrate that undecidability is not just an abstract notion limited to theoretical computer science, but a relevant and thought-provoking concept that has tangible implications for contemporary music composition and production.
\subsection{Undecidability results}
We will achieve this by investigating four distinct areas of music making:
\begin{enumerate}
    \item Proving that Ableton, a popular digital audio workstation, is Turing complete.
    \item Establishing the undecidability of satisfiability in sound creation using audio effects.
    \item Demonstrating the undecidability of constraints on polymeters in musical compositions.
    \item Proving that the satisfiability of just intonation harmony constraints is undecidable.
    \item Proving that the satisfiability of "new ordering systems" is undecidable.
\end{enumerate}

These will all rely on reductions to problems already known to be undecidable in the computer science literature.
\subsection{Relevance of these results}
We will also provide detailed explanations of why someone who is not a theoretical computer scientist should care about each of these properties.  In particular, we will discuss why Rice's theorem argues for a real limitation in the possibility of analyzing output of a Turing complete system (which Ableton is), and that undecidability of satisfiability constraints provides a theoretical limit on to what extent an (automated or human) composer can "think abstractly, then compose concretely" within different domains.

\section{Decidability, Turing Completeness, and Their Implications for Music Production}

This section aims to introduce the fundamental concepts that underpin our exploration of Ableton Live's computational capabilities and the implications of these for the undecidability of general properties of Ableton projects and the satisfiability of a composer's constraints or vision.

\subsection{Decidability: An Overview}

Decidability, a concept from theoretical computer science, refers to the solvability of a problem through algorithmic means. A problem is considered decidable if an algorithm exists that can provide a definitive solution to every instance of the problem in a finite timeframe. Conversely, a problem is undecidable if no such algorithm can be found. Understanding decidability is crucial as it delineates the boundary between problems that can be addressed using computational techniques and those that remain fundamentally unsolvable.

\subsection{Turing Machines and Turing Completeness}

The concept of a Turing machine, a theoretical model of computation, is central to understanding the abilities and limitations of computational systems. A system or programming language is considered Turing complete if it can simulate a Turing machine's behavior, meaning that it can execute any computation that a Turing machine can, given adequate time and resources.

Ableton Live, a digital audio workstation, is such a Turing complete system. Its computational power, coupled with its capability to manipulate audio and MIDI data, makes it an immensely flexible tool for musical creation. However, this Turing completeness also implies that certain questions about the behavior of Ableton Live projects are undecidable, leading to intriguing implications for music production.

\subsection{Rice's Theorem and Its Consequences}

Rice's theorem is a result in computability theory that states any non-trivial property concerning a Turing machine's computed function is undecidable \cite{rice}. This theorem suggests that many questions about the behavior of computational systems and programs are fundamentally unanswerable, including those related to the properties of Ableton Live projects.

Thus, the power and flexibility of Ableton Live, as embodied in its Turing completeness, bring with them a fascinating paradox. While they allow for virtually limitless musical creativity, they also introduce elements of undecidability that challenge our ability to fully understand or predict the outcomes of complex musical projects.

\subsection{Undecidability of Satisfiability Problems}

Satisfiability problems, which involve determining whether there exists a solution that satisfies a set of conditions or constraints \cite{sat}, also encounter undecidability issues. These problems are pervasive in computer science and mathematics, often manifesting in various domains, including logic \cite{logic}, number theory \cite{numbertheory} , and algebra \cite{algebra}.

In the context of music composition, satisfiability problems can occur when a composer defines a specific "vision" for their piece, expressed in terms of rhythmic, harmonic, or structural constraints. This vision might involve intricate polyrhythms, complex harmonic progressions, or specific structural properties.

However, the satisfiability of these constraints—whether there exists a musical piece that fulfills all of them—is often undecidable. This means there is no algorithm that can determine, for every possible set of constraints, whether a satisfying musical piece exists. Therefore, a composer, whether human or machine, might have a vision for a piece without any guarantee that it can be realized.

In the following sections, we will explore how these undecidability issues can limit a composer's ability to know a priori if their vision can ever be realized, highlighting the inherent tension between the vast expressive power of a Turing complete system like Ableton Live and the fundamental limits imposed by undecidability.

\section{Ableton Live's Turing Completeness}
\subsection{An Introduction to Ableton Live}
Ableton Live is a comprehensive digital audio workstation (DAW) that offers a diverse set of tools for computer musicians, catering to various aspects of music production and performance. A key feature of Ableton Live is its flexible audio and MIDI routing capabilities. You can freely route audio and MIDI between tracks, enabling complex signal flows, layering of sounds, and intricate cross-processing. This flexibility opens up vast creative possibilities, from creating intricate soundscapes to designing complex rhythmic patterns.

Additionally, Ableton Live provides a vast array of MIDI and audio effect devices. MIDI effects transform MIDI notes and control signals, influencing parameters such as pitch, velocity, and timing. Audio effects manipulate the sound, offering control over parameters such as frequency, amplitude, and time-based effects. What's even more compelling is that these effects can interact with each other, allowing the parameters of one effect to influence another. This interoperability allows for the crafting of unique sonic textures and innovative musical ideas. The combination of these features makes Ableton Live a versatile and powerful tool, enabling musicians to push boundaries and expand their creative potential in music production and performance.
\subsection{Proof of Turing Completeness}
In this section, we present our proof of Ableton Live's Turing completeness by constructing a Turing machine simulation using its built-in audio and MIDI devices.

\subsubsection{Infinite Tape}
Create an audio track representing the infinite tape with an unbounded audio recording. Each audio sample (or short sequence of samples) represents a symbol, mapped to different frequency ranges.

\subsubsection{Read/Write Head}
\begin{itemize}
\item Reading: Use Granulator II to read the audio track in real-time. Automate the FilePos parameter to control the read head's position. Use EQ Eight to isolate the frequency range representing a specific symbol, and use Envelope Follower to analyze the output, determining the symbol at the current position.

\item Writing: Use Utility to control the amplitude of the audio track. Add an Expression Control device and map its output to the Gain parameter of the Utility device, effectively overwriting the symbol at the current position with the new symbol according to the Turing machine's rules.
\end{itemize}
\subsubsection{States}
Place an Audio Effect Rack on the reading track, containing multiple audio effect chains, each representing a different state of the Turing machine. Each chain contains a combination of audio and MIDI effects that determine the next state and tape action based on the symbol read by Granulator II and analyzed by the Envelope Follower.

\subsubsection{Rules}
Implement the Turing machine's rules using MIDI Effect Racks, Chord devices, and MIDI routing in this manner

\begin{itemize}
\item Create an audio track for the infinite tape and read/write head, referred to as the "reading track." This track will contain the Granulator II device for reading symbols, the EQ Eight and Envelope Follower for symbol analysis, and the Utility and Expression Control devices for overwriting symbols.
\item Create a separate MIDI track for each state in the Turing machine. Label these tracks as "State 1," "State 2," and so on.
\item Route the MIDI output of each "State" track to the Chain Selector parameter of the Audio Effect Rack on the reading track. To do this, set the "MIDI To" option in the I/O section of each "State" track to the reading track, and then select "Chain Selector" as the target parameter.
\item On each "State" MIDI track, place a MIDI Effect Rack with multiple chains, each chain representing a different rule for the current state based on the input symbol. For example, in "State 1" track's MIDI Effect Rack, create chains for each possible symbol the read/write head may encounter while in State 1.
\item Configure the Chain Selector in the MIDI Effect Rack on each "State" track to choose the appropriate chain based on the symbol read by the read/write head. Use the Envelope Follower's output value (from the reading track) to control the Chain Selector on the corresponding "State" track. To do this, you can use MIDI mapping or automation.
\end{itemize}

\subsection{The Validity of the Proof} The method we employed to prove Ableton Live's Turing completeness is a widely accepted technique within the realm of theoretical computer science \cite{reduction}. It's based on the concept of simulation, where the key components of a Turing machine - an infinite tape, a read/write head, states, and transition rules - are mapped to corresponding functionalities within Ableton Live's music production environment. By demonstrating that Ableton Live can effectively simulate a Turing machine, we establish its Turing completeness. This is because any system that can simulate a Turing machine, which is the theoretical model for all computation, is itself Turing complete. Therefore, our mapping approach not only demonstrates the richness and flexibility of Ableton Live as a music production tool but also highlights its computational universality.

\subsection{Implications of Turing completeness in music production}
One implication of Ableton’s Turing completeness is that it is, in theory, possible to build any program in Ableton (a neural network, a compiler of a different equation, etc.)  However, I argue that the most important implication of Ableton’s Turing completeness for music analysts and creators (particularly those interested in automating musical tasks) comes from Rice’s theorem.  Imagine that we are designing an audio installation that is supposed to go on indefinitely, and we want to avoid ever having the perception of dissonance (defined according to an existing metric like having lots of energy at two beating frequencies).  According to Rice’s theorem, there does not exist a program in any language which, regardless of the contents of the installation, could check that it observes that property.  Similarly, if an automated system wanted to design  procedural music to a videogame which continued as long as the game was being played and responded to the game’s inputs, there could never be a program that could check every such procedural music and determine whether it responds in a logical way.  This undecidability result means that by using Ableton’s complexity, we must sacrifice some degree of certainty about its outputs.

\section{The Undecidability of Understanding Audio Effects}

Consider a scenario where a sound engineer aspires to generate a specific audio output using a particular audio effect. This effect can be modeled as a real-valued function. The question that might arise in such a scenario is: "Given any possible input audio signals, will applying this specific audio effect consistently produce the desired output audio signal?"

To figure out whether we can answer this, we must turn to work in computability theory.  The undecidability of the universal theory of the reals, a result that stems from the seminal work of mathematicians like Alonzo Church, Alan Turing, and others, is a profound and far-reaching theorem in mathematical logic and computability theory \cite{church} \cite{turing1} \cite{turing2}. It states that there does not exist a universal algorithm that can decide whether an arbitrary mathematical statement concerning real numbers is true or false. This result is a consequence of Gödel's incompleteness theorem and the negative solution to Hilbert's Entscheidungsproblem, the decision problem.  Note that certain statements about reals, e.g., first order theories of real numbers, are decidable \cite{tarski}; however, they become undecidable when the sine function (which is fundamental to audio processing) is included in the theory.

Now, let's translate this abstract mathematical concept into our audio engineering context. The real-valued function representing an audio effect can be thought of as an equation involving real numbers. Just like the equations in the universal theory of the reals, this audio effect function might have many possible inputs, which correspond to the different potential input audio signals. The output of this function, meanwhile, corresponds to the resulting sound.

When the sound engineer seeks to find an audio effect that can consistently produce a desired output from any possible input, they are essentially trying to solve a problem similar to deciding whether a specific equation holds for all real numbers. In mathematical terms, they are trying to find a function (audio effect) that satisfies a particular condition (produces the desired sound) for all possible inputs (audio signals).  Furthermore, they are likely implicitly including functions such as the sine function in their equation.

Given the undecidability result of the universal theory of the reals, however, we know that there cannot exist a universal algorithm capable of deciding whether an arbitrary equation holds for all real numbers. By analogy, then, there cannot exist a universal algorithm capable of deciding whether an arbitrary audio effect can produce a desired output from any possible input.

This has far-reaching implications for sound engineering and, more broadly, any field that involves transformations of real-valued signals or data. It suggests that there may be no systematic method for determining whether a particular transformation (such as an audio effect) can achieve a specific desired result from any possible input. 

For a human music producer, the undecidability of understanding audio effects can have significant implications. On the one hand, it emphasizes the limits of formal, algorithmic approaches and the potential unpredictability of audio production processes. This can encourage producers to embrace a more exploratory, creative, and intuitive approach to using audio effects, leveraging their personal experiences, aesthetic judgments, and experimental practices. On the other hand, it also highlights the inherent complexity and open-ended nature of audio production, which can be both challenging and exciting. It suggests that there are no definitive answers or universal recipes for achieving specific sounds, and that the creative possibilities are vast and potentially unbounded.

In the case of automated systems for music production, the undecidability of understanding audio effects can also have important consequences. It implies that there are inherent limitations to the capabilities of these systems, particularly in terms of their ability to predict and control the outcomes of applying audio effects. This can impact the design and development of such systems, emphasizing the need for robustness, adaptability, and flexibility. It might also necessitate the use of probabilistic and heuristic methods, as well as machine learning techniques, which can cope with uncertainty and make educated guesses in the face of undecidability. However, it also opens up opportunities for creative applications of AI in music production, where the unpredictability of audio effects is not a bug, but a feature that can be used to generate new and unexpected musical ideas.

\section{Undecidability of Satisfiability with Polyrhythms or Just Intonation}
\subsection{Undecidability of Satisfiability with Multiplication of Unknown Integers}
The theory of undecidability of integer variable multiplication traces its origins to the field of mathematical logic and theoretical computer science. Here, it is associated with the principle of the universal theory of the reals, which focuses on the properties of real numbers and the mathematical structures derived from that set. This theory consists of several profound results, including the undecidability of multiplication involving unknown integer variables. 

The undecidability of the satisfiability of multiplication by unknown integer variables is not an obvious concept, and it took revolutionary work by computer scientists and logicians to bring it to light. Essentially, the claim of this theory is that there is no universal method or computational process (termed 'algorithm') that can definitively determine whether any general mathematical statement, particularly one involving the multiplication of real numbers or unknown variables, is always true or false under all circumstances. 

This proposition was explicitly proved through the work of Yuri Matiyasevich, building on contributions by three leading logicians of the 20th century: Julia Robinson, Martin Davis, and Hilary Putnam \cite{davis} \cite{robinson} \cite{putnam}. Their quest to resolve a significant mathematical problem of the time, known as Hilbert's Tenth Problem, inadvertently led to the discovery of the undecidability. Hilbert's Tenth Problem asked for an algorithm that could determine whether any given Diophantine equation had integer solutions. A Diophantine equation is a polynomial equation that seeks integer solutions.

Matiyasevich built upon the work of the aforementioned logicians, and in 1970, he provided the final piece of the puzzle, proving that no such algorithm exists for Hilbert's Tenth Problem, thus confirming the resolution of the problem in the negative \cite{matisayevitch}. This negative resolution subsequently implied that certain specific sets, in this case, the sets that represent the satisfiability of certain Diophantine equations or equivalently, the multiplication of integer unknown variables, are undecidable. 

\subsection{Relationship to Polymeter and Polyrhythm}

In metric music, beats can be visually represented using pairs of integers, labeled as (n, m). Here, 'n' signifies the beat number or the position of the beat within a rhythmic cycle and 'm' represents the time signature. This notation system enables the mapping of complex rhythms onto a numeric grid and permits us to vislauize and grapple with abstract musical concepts. 

Taking a hypothetical scenario in which we want two beats from different rhythms to coincide at a definite point in time, we can denote this intention using an equation: (n1, k1) * (n2, k2) = (a, b). Here, (n1, k1) and (n2, k2) correspond to the starting beats of the two rhythms, while (a, b) symbolizes the point of conjunction. The variables k1 and k2 in this equation are unknown, thereby making the equation a case of multiplication involving unknown variables.

The theory of the undecidability of the satisfiability of multiplication by unknown variables becomes highly relevant here. There are no definitive computational procedures that can yield actual values for k1 and k2, making the equation true. This implies that it's not systematically determinable, on solely mathematical grounds, whether we can make two beats from differing rhythms coincide at a specific time point—a given polyrhythm feasible.

This undecidability offers profound implications for music theory and composition, particularly when it comes to polyrhythms. Composers often have to deal with the complexity and unpredictability that comes with writing polyrhythms. The challenge lies in systematically specifying and reasoning about whether it is possible for certain beats within a complex polyrhythm to align. Because the algorithmic and procedural methods cannot provide definitive answers due to the undecidability at play, composers must often resort to heuristic or empirical methods. These methods, based on experimentation, trial and error, or intuitive judgment, highlight the often underappreciated complexity and creative challenge in music composition.
\subsection{Just Intonation}
Just intonation is a system of musical tuning in which the frequencies of notes are related by ratios of small whole numbers. Theoretically, it results in a pure and consonant sound that is often more pleasing to the ear compared to other tuning systems. However, the process of creating just intonation involves the multiplication of ratios, which inherently corresponds to the multiplication of unknowns in the context of mathematical equations.

The undecidability of the satisfiability of multiplication by unknowns directly impacts our capacity to devise a universally applicable algorithm for creating just intonation automatically. In other words, there is no general algorithm or decision procedure that can definitively determine whether it's possible to tune a piece of music in just intonation by multiplying specific frequency ratios. This undecidability makes it inherently complex to systematize the process of creating just intonation.

For instance, let's consider a situation where we want to tune a piece of music that uses a seven-note diatonic scale in just intonation. We may want to determine the frequency ratios that will provide a perfect fifth (a frequency ratio of 3:2) between each pair of successive notes. We could express this requirement using equations involving multiplication of unknowns, such as a/b * c/d = 3/2, where a/b and c/d represent the frequency ratios of the two notes. However, due to the undecidability of the satisfiability of multiplication by unknowns, we can't conclusively determine algorithmically whether there exist specific values of a, b, c, and d that will satisfy these equations for all pairs of notes. 
.
\subsection{Implications}
For human composers, the undecidability of these harmonic and rhythmic constraints has profound implications. It means that there might be times when they conceive of a musical idea, a particular harmonic or rhythmic language, but they might not be able to determine if it is even plausible to realize that idea given the constraints of the harmonic space. The inability to formally or algorithmically reason about these constraints can introduce an element of profound uncertainty into the creative process. Rather than seeing this as a limitation, it can be interpreted as an invitation to the composer to delve deeper into the rich and unexplored territories of the harmonic space. This uncertainty, then, serves not as a boundary, but as a stimulus for innovation and experimentation in music composition.

When it comes to machine models used for automatic composition or music analysis, the undecidability of reasoning about just intonation harmony or polyrhythms presents substantial challenges as well. It suggests that AI systems cannot be guaranteed to always find a solution to satisfy certain harmonic constraints, or even determine whether such a solution exists. This necessitates the development of AI models that are capable of handling these inherent uncertainties, perhaps by leveraging probabilistic models or machine learning techniques. At the same time, the undecidability could also be seen as a source of creative potential, enabling AI models to generate novel and unexpected musical ideas within the uncharted territories of the harmonic space.

\subsection{The Undecidability of Horn Clauses and Composer-Generated Systems of Organization}
\subsection{Proof of undecidability of Horn Clauses}
The proof for the undecidability of Horn clauses is rooted in the reduction from the Halting Problem, which is a well-known undecidable problem within computer science. One begins by establishing that the set of all true ground (i.e., variable-free) Horn clauses is recursively enumerable \cite{apt1988towards}. This means that there exists a Turing machine which will list all true Horn clauses. Then, a reduction is crafted from the Halting Problem to the problem of deciding if a given ground Horn clause is true. The subsistence of this reduction implies that if there were an algorithm to decide the truthfulness or groundness of any given Horn clause, such an algorithm could be used to solve the Halting Problem, contradicting its established status as undecidable. 

David A. Plaisted, in his comprehensive study "The Undecidability of Ground Term Rewrite Systems" \cite{plaisted}, was able to prove that establishing satisfiability for sets of Horn clauses is undecidable when clauses could contain function symbols of arity $>0$. His work is considered significant in demonstrating the undecidability of Horn clauses satisfiability. Plaisted obtained this result by reducing the word problem for semi-Thue systems (which is a known undecidable problem) to the satisfiability problem of Horn clauses.

\subsection{Implications for Composers}
The notion of equating a musical system to general Horn clauses  suggests that each musical rule in a system, whether it governs melody, harmony, or rhythm, can be formulated as a Horn clause. 

In order to achieve this transmutation, one must assign a literal or proposition to every possible musical event or condition. For instance, we can designate 'note X is followed by note Y' as a positive literal, and 'chord progression A leads to chord progression B' as a negative literal or a combination of both. As such, each literal serves as a concrete representation of a musical condition. These literals are the building blocks with which the Horn clauses are formulated, effectively creating a musical language with syntactic rules and structures mirroring those in propositional logic.

Collectively, these translated musical rules defined by a composer form a system that resembles the characteristics of a set of general Horn clauses in propositional logic. Each rule represented as a Horn clause becomes a stipulation that demands satisfaction. In this context, satisfaction implies a musical sequence that corresponds and adheres to the designed compositional system or the logical constraints impounding it.

In propositional logic, the notion of whether a model exists that satisfies a given set of clauses is described as the satisfiability problem. Analogously applied to the domain of music, this translates to the existential question: can a sequence of musical events, a composition, be constructed satisfying all the rules of a given system simultaneously? 

However, drawing this parallel presents an inherent challenge grounded in the conclusively proven undecidability of general Horn clause satisfiability. This paradigm transforms from an intellectual exercise into a practical dilemma for composers who wish to implement such systems in their work. Without guaranteed certifying mechanisms, composers following this approach walk a delicate path between logical structure and potential unsatisfiability. The pursuit of an aesthetically pleasing composition that simultaneously adheres to intricate logical constraints places an extraordinary demand on both the creative and logical faculties of the composer.
\section{Conclusion}

In this paper, we have presented several key findings that have profound implications for both the practice of music composition and the development of automated music generation systems. Our results highlight the inherent complexities and limitations encountered when attempting to computationally represent and generate musical pieces, whether the context is traditional musical forms like string quartets or more experimental formats such as real-time, non-time-bounded media installations.

Our first major result was establishing the Turing completeness of Ableton Live, a popular digital audio workstation. This result not only underscores Ableton Live's vast expressive power as a musical tool but also sets the stage for our subsequent investigations into the inherent computational limits of music creation within such a powerful system.

We then demonstrated the undecidability of satisfiability for polyrhythmic and just intonation constraints, which are common elements in contemporary music composition. This result underscores the fundamental limits of formalizing and algorithmically processing certain aspects of musical composition. We showed that even with a Turing complete system, there is no guarantee that a composer's specific vision for a piece can always be realized, whether the constraints involve intricate rhythmic structures, complex harmonic progressions, or other specific musical properties.

Moreover, we extended this result to general composer-generated structural constraints, further highlighting the challenges composers face when attempting to realize their musical ideas, particularly when those ideas involve intricate or complex structural properties.

These findings have significant implications for both human and automated composers. For human composers, our results underscore the importance of intuition, experience, and exploratory techniques in the composition process, particularly when dealing with complex musical constraints that cannot be fully formalized or algorithmically processed. For automated systems, our results highlight the limitations of relying solely on algorithmic methods for generating music that satisfies specific constraints. They also underscore the need for incorporating heuristic or learning-based methods to navigate the vast and complex landscape of musical possibilities.

In conclusion, our findings serve as a bridge between the worlds of computational theory and music composition, shedding light on the fascinating interplay between computation and creativity, and ultimately, between machines and the art of music.
\printbibliography
\end{document}